\documentclass[twocolumn,amsmath,amssymb,superscriptaddress,prb,floatfix]{revtex4}

\usepackage{graphicx}% Include figure files
\usepackage{dcolumn}% Align table columns on decimal point
\usepackage{bm}% bold math
\usepackage[dvips,dvipsnames,usenames]{color}

\def\R{{\bf R}}

\newcommand{\ket}    [1]{{|#1\rangle}}
\newcommand{\bra}    [1]{{\langle#1|}}
\newcommand{\braket} [2]{{\langle#1|#2\rangle}}
\newcommand{\bracket}[3]{{\langle#1|#2|#3\rangle}}

\begin{document}

\title{A First Principles Theory of Nuclear Magnetic Resonance
  J-Coupling in solid-state systems.}

\author{Si\^{a}n A. Joyce}
\affiliation{Tyndall National Institute, Lee Maltings, Prospect Row, Cork, Ireland} 
\author{Jonathan R. Yates}
\affiliation{TCM Group, Cavendish Laboratory, University of Cambridge, UK}
\author{Chris J. Pickard}
\affiliation{School of Physics and Astronomy, University of St. Andrews, St. Andrews, Scotland}
\author{Francesco Mauri}
 \affiliation{Institut de Min\'{e}ralogie et de Physique des Milieux Condens\'{e}s,
Universit\'{e} Pierre et Marie Curie, Paris, France}

\date{\today{}}

\begin{abstract}
A method to calculate NMR J-coupling constants from first principles
in extended systems is presented. It is based on density functional theory and is
formulated within a planewave-pseudopotential framework. The
all-electron properties are recovered using the
projector augmented wave approach. The method is validated by
comparison with existing quantum chemical calculations of 
solution-state systems and with experimental data. The approach has been applied to 
verify measured J-coupling in a silicophosphate structure, Si$_{5}$O(PO$_{4}$)$_{6}$.
\end{abstract}

\maketitle

%%%%%%%%%%%%%%%%%%%%%%%%%% Intro %%%%%%%%%%%%%%%%%%%%%%%%%%%
\section{Introduction}

Nuclear Magnetic Resonance (NMR) allows information to be relayed
through magnetic nuclei in a non-destructive and powerful
approach to structural elucidation. It is a fundamental tool in a 
broad range of scientific disciplines and is a cornerstone 
of modern spectroscopy. NMR spectra yield a wealth of information, 
the most commonly reported property being the chemical shift.
This parameter relates an externally applied 
magnetic field to the resulting change in the local electronic 
environment of the magnetic nuclei, thereby providing key insight 
into the underlying atomic structure.

NMR J-coupling or indirect nuclear spin-spin coupling is an 
indirect interaction of the nuclear magnetic moments 
mediated by the bonding electrons. It is manifested as the
fine-structure in NMR spectra, providing a
direct measure of bond-strength and a map of the connectivities of the
system. The J-coupling mechanism is an essential component of many 
NMR experiments.\cite{levitt} 

In solution-state, J-coupling measurements can
often be obtained from one dimensional spectra where the multiplet splitting in the
peaks is clearly resolved. However, in the solid-state this is not the
case as these splittings are typically obscured by the broadenings 
from anisotropic interactions. Fortunately this technical challenge 
has not prevented the determination of J-coupling in the solid-state, as recent work
employing spin-echo Magic Angle Spinning (MAS) techniques\cite{duma04} has resulted
in accurate measurements of J-coupling in both inorganic
materials\cite{amoureux05,coelho06,cadars07, coelho07} and molecular crystals.\cite{lai06,
 brown04,brown02,brown02b,pham07} In combination with the advances in
solid-state experiments, there has also been an increased interest 
from the biomolecular community as J-coupling has been found to be a
direct measure of hydrogen bond strength.\cite{dingley98,dingley05, pham07}
Both of these factors have provided a strong impetus to develop first principles
approaches to compute the NMR J-coupling constants in order to 
support experimental work, particularly for solid-state systems.

For finite systems, NMR parameters, including both chemical shifts and J-couplings, can be
routinely calculated using traditional quantum chemistry approaches based
on localised orbitals.\cite{vaara02, helgaker99} 
Such calculations have been widely applied to 
 assign the solution-state NMR spectra  of  molecular systems and establish key
conformational and structural trends.\cite{nmr_book} In particular NMR J-couplings have
been used to quantify hydrogen bonding\cite{grzesiek04} in biological systems.
In  order to apply these techniques to solid-state NMR, it is necessary
to devise finite clusters of atoms which model the local environment
around a site of interest in the true extended structure.  While this 
 has led to successful studies of NMR chemical shifts in systems such as molecular
crystals,\cite{facelli93} supra-molecular assemblies\cite{ochsenfeld01} and organo-metallic 
compounds,\cite{salzmann98} it is clear
that there are advantages in an approach that inherently takes account of
the long-range electrostatic effects in extended systems.

 This observation has led to the recent development of  the  Gauge 
Including  Projector  Augmented  Wave  (GIPAW)\cite{pickard01}  method  which  enables   NMR
parameters to be calculated at all-electron accuracy within the
planewave-pseudopotential formalism\cite{payne1} of density functional theory (DFT). 
 The  technique  has  been  applied,  in  combination  with
experimental  NMR  spectroscopy,  to  systems  such  as
minerals,\cite{profeta03,ashbrook06,farnan03} 
glasses\cite{benoit05,charpentier04}   and   molecular
crystals.\cite{yates04,yates05-malt,gervais05} 

In this paper we introduce a theory to compute NMR J-Couplings 
in extended systems using periodic boundary conditions and supercells
with the planewave-pseudopotential approach. Like the GIPAW approach to calculating
NMR chemical shifts, our method is formulated within the planewave-pseudopotential framework
using density functional perturbation theory (DFPT). We use the
projector-augmented-wave\cite{blochl1} (PAW) 
reconstruction technique to calculate J-couplings with all-electron accuracy. 

In the following section we discuss the physical mechanism of the indirect
spin-spin interaction, the basis of the PAW approach and the supercell
technique.  In Sections
 \ref{sec:mag} and \ref{sec:cur} we show how the J-coupling tensor maybe be calculated using
 PAW and DFPT. The method has been implemented in a parallel plane-wave electronic
structure code and we discuss details of the implementation and provide
validation results in Section \ref{sec:res}.

%%%%%%%%%%%%%%%%%%%%%%%%% Theory %%%%%%%%%%%%%%%%%%%%%%%%%%%
%INTRO
\section{NMR J-Coupling}\label{sec:intro2}
We consider the interaction of two nuclei, ${\rm K}$ and
${\rm L}$, with magnetic moments, ${\bm \mu}_{{\rm K}}$ and ${\bm \mu}_{{\rm
    L}}$, mediated by the electrons. The first complete analysis of this indirect coupling was
provided by Ramsey\cite{ramsey52,ramsey53} who decomposed the interaction into four distinct
physical mechanisms; two involving the interaction of the nuclear spins
through the electron spin and two through the electron charge. In the absence of
spin-orbit coupling i.e, for relatively light elements, the charge and
spin interactions can be treated separately. 
We can write the magnetic field at atom ${\rm L}$ induced
by the magnetic moment of atom ${\rm K}$ as
\begin{eqnarray}\label{eq:b_ind}
{\bf B}^{(1)}_{\rm in}({\bf R}_{{\rm L}}) & = & \frac{\mu_{0}}{4\pi}\int 
\left[\frac{3({\bf m}^{(1)}({\bf r})\cdot {\bf r}_{{\rm L}}){\bf r}_{{\rm L}} - 
{\bf m}^{(1)}({\bf r})|{\bf r}_{{\rm L}}|^{2}}{|{\bf r}_{{\rm L}}|^{5}}\right]\,{\rm d}^{3}{\bf r} \nonumber \\
                          & + & \frac{\mu_{0}}{4\pi}\frac{8\pi}{3}\int {\bf m}^{(1)}({\bf r})
\delta({\bf r}_{{\rm L}})\,{\rm d}^{3}{\bf r} \nonumber \\
                          & + & \frac{\mu_{0}}{4\pi}\int {\bf j}^{(1)}({\bf r})\times
\frac{{\bf r}_{{\rm L}}}{|{\bf r}_{{\rm L}}|^{3}}\,{\rm d}^{3}{\bf r}.
\end{eqnarray}
${\bf r}_{{\rm L}} = {\bf R}_{{\rm L}} - {\bf r}$, where ${\bf R}_{{\rm L}}$ is the
position of nucleus ${\rm L}$, $\mu_{0}$ is the permeability 
of a vacuum and $\delta$ is the Dirac delta function. 

${\bm \mu}_{{\rm K}}$ interacts with the electron spin 
through a magnetic field generated by a 
Fermi-contact term, which is due to the finite probability of the
presence of an electron at the nucleus, and a spin-dipolar interaction. 
Both of these terms give rise to a first order spin
magnetisation density, ${\bf m}^{(1)}({\bf r})$. This magnetisation
density then induces a magnetic field at the receiving nucleus by
the same mechanisms, which in this case are given respectively by the first
and second terms of Eqn.~\ref{eq:b_ind}.
The interaction between $\bm{\mu}_{{\rm K}}$ and the electronic charge
gives rise to an induced current density.  
To first-order this is given by ${\bf j}^{(1)}({\bf r})$ and can be divided
into a paramagnetic and a diamagnetic contribution. 

 The J-coupling tensor,
$\overleftrightarrow{\bf J}_{{\rm LK}}$, between ${\rm L}$ and ${\rm K}$, can be related
to the induced field by
\begin{equation}\label{eq:J}
{\bf B}^{(1)}_{\rm in}({\bf R}_{{\rm L}})  = \frac{2\pi}{\hbar\gamma_{{\rm L}}\gamma_{{\rm K}}}{\overleftrightarrow{\bf
    J}}_{{\rm LK}} \cdot {\bm \mu}_{{\rm K}}, 
\end{equation}  
where $\gamma_{{\rm L}}$ and $\gamma_{{\rm K}}$ are the gyromagnetic ratios of
nuclei ${\rm L}$ and ${\rm K}$. Although the physical 
interpretation of J-coupling is simplified by considering the interaction in
terms of a responding and a perturbing nucleus, it is a symmetric coupling and 
either atom ${\rm L}$ or ${\rm K}$ can be considered as the perturbing site.
Experimental interest is focused primarily on the isotropic coupling 
constant, ${\rm J}^{n}_{{\rm LK}}$, which is obtained from the 
trace of ${\overleftrightarrow{\bf J}}_{{\rm LK}}$ and is measured in
Hz. The superscript, $n$, denotes the order of the coupling in terms
of the number of bonds separating the coupled nuclei. In a typical NMR
experiment J-coupling can be measured across a maximum of three bonds.\cite{marquez01,edden04}
In this paper we concentrate solely on obtaining the isotropic or scalar value. 

To calculate ${\overleftrightarrow{\bf J}_{{\rm KL}}}$ we obtain ${\bf
  m}^{(1)}$ and ${\bf j}^{(1)}$ within density functional perturbation 
theory using a planewave expansion for the wavefunctions with
periodic boundary conditions and pseudopotentials to represent the ionic
  cores. 
The use of pseudopotentials generates a complication as the
 J-coupling tensor depends critically on the wavefunction in the
  regions close to the perturbing and receiving nuclei, precisely the
  regions where the pseudo-wavefunctions have a non-physical form.
To compensate for this 
we perform an all-electron reconstruction of the valence wavefunctions
  in the core region using Bl\"{o}chl's projector augmented wave (PAW) scheme.\cite{walle1} 

Within this scheme, the expectation value of an operator $O$, applied to
the all-electron wavefunctions, $\ket{\psi}$, is expressed in terms of
the pseudised wavefunctions, $\ket{\widetilde{\psi}}$, as:
$\bracket{\psi}{O}{\psi} =
\bracket{\widetilde{\psi}}{\widetilde{O}}{\widetilde{\psi}}$. Here,
for an all-electron local or semi-local operator, $O$, the
corresponding pseudo-operator, $\widetilde{O}$, is given by
\begin{eqnarray}\label{eq:paw}
\widetilde{O}&  = & O + \sum_{{\bf R},n,m}\ket{\widetilde{p}_{{\bf
      R},n}}\left[\bracket{\phi_{{\bf R},n}}{O}{\phi_{{\bf R},m}}\right. \nonumber \\
          &  & -  \bracket{\widetilde{\phi}_{{\bf R},n}}{O}{\widetilde{\phi}_{{\bf
      R},m}}\left.\right]\bra{\widetilde{p}_{{\bf R},m}}. 
\end{eqnarray}
${\bf R}$ labels the atomic site, or augmentation region, and $n$ and 
$m$ are composite indexes which account for the angular momentum
channels and the number of projectors.
$|\phi_{{\bf R},n}\rangle$ are the all-electron partial waves obtained
as eigenstates of an atomic calculation within $r_{c}$, the 
pseudopotential core radius and $|\widetilde{\phi}_{{\bf R},n}\rangle$ are
the corresponding pseudo partial waves. $|\widetilde{p}_{{\bf
    R},n}\rangle$ 
are the localised projectors which weight the superposition of partial 
waves where $\langle\widetilde{p}_{{\bf R},n}|\widetilde{\phi}_{{\bf R}',m}\rangle =
\delta_{{\bf RR}'}\delta_{nm}$. The PAW method has been
used to calculate several all-electron properties from pseudopotential
calculations including: EPR hyperfine parameters,\cite{walle1} 
electric field gradient tensors\cite{efg1} and  Electron
Energy Loss Spectroscopy.\cite{eels1} 

\begin{figure}
\includegraphics*[width=8.0cm]{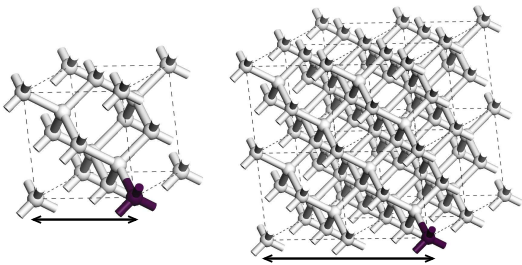}
\caption{\label{fig:supercell}Schematic of the supercell
  technique. The unit cell is on the left, a 2$\times$2$\times$2 supercell of the
  unit cell is on the right. This supercell doubles the distance
  between the perturbing atom (black) and its periodic image in the
  next cell.}
\end{figure}

To calculate J-couplings in the solid-state using periodic boundary
conditions, the perturbing nucleus can be viewed similar to a defect
in a defect calculation. This allows us to use the 
standard technique of constructing supercells from the unit cell which
are large enough to inhibit the interaction between the periodic defects or 
perturbations. This corresponds 
to extending the system-size to facilitate the decay of the induced
magnetisation and current densities within the simulation cell. 
Figure. \ref{fig:supercell} is a schematic of a
$2\times2\times2$ supercell constructed from eight unit cells. The
perturbing atom now lies at the corner of a much larger cell which
decreases the interaction between the perturbation and its periodic
image. This approach works very well for localised properties such as
J-coupling. To calculate the J-coupling for molecules, we use a vacuum supercell 
technique. In both cases, the J-couplings must be converged with respect
to the cell-size. 

%%%%%%%%%%%%%%%%%%%%%%%%%%%%%%%%%%%%%%%%
\section{The Spin Magnetisation Density}\label{sec:mag}
%%%%%%%%%%%%%%%%%%%%%%%%%%%%%%%%%%%%%%%%
We now obtain the contribution to the J-coupling tensor which arises from the
interaction of the nuclear spins mediated by the electron spin.
We first obtain an expression for the pseudo-Hamiltonian in the 
presence of a perturbing nuclear spin and show how it can be used to 
obtain the induced magnetisation density. We then use the magnetisation 
density to calculate the magnetic field induced at the receiving nucleus.
%%%%%%%%%%%%%%%%%%%%%%%%%%%%%%%
\subsection{Pseudo-Hamiltonian}
%%%%%%%%%%%%%%%%%%%%%%%%%%%%%%%
The all-electron Hamiltonian for a system containing ${\rm N}$ magnetic
moments which interact through the electron spin, ${\bf S}$, is expanded
to first order in the magnetic moment of the
perturbing site, ${\bm \mu}_{{\rm K}}$, to give
\begin{equation}\label{eq:ae-1st}
{\rm H} = \frac{1}{2}{\bf p}^{2} + {\rm V}^{(0)} ({\bf r}) + {\rm
  V}^{(1)}({\bf r})
 + {\rm H}_{\rm SD} + {\rm H}_{\rm FC}. 
\end{equation}
where
\begin{equation}
{\rm H}_{\rm SD}=g\beta{\bf S}\cdot {\bf B}_{{\rm K}}^{\rm SD},
\end{equation}
and
\begin{equation}
{\rm H}_{\rm FC}=g\beta{\bf S}\cdot {\bf B}_{{\rm K}}^{\rm FC}.
\end{equation}
${\bf B}_{{\rm K}}^{\rm SD}$ is the magnetic field generated by a spin-dipole interaction,
\begin{equation}
{\bf B}_{{\rm K}}^{{\rm SD}}  = \frac{\mu_{0}}{4\pi}\frac{3{\bf r}_{{\rm K}}({\bf r}_{{\rm K}}
\cdot{\bm \mu}_{{\rm K}}) - r_{{\rm K}}^{2}{\bm \mu}_{{\rm K}}{\cal I}}{|r_{{\rm K}}|^{5}}, 
\end{equation}
and ${\bf B}_{{\rm K}}^{\rm FC}$ is the Fermi contact interaction,
\begin{equation}
{\bf B}_{{\rm K}}^{{\rm FC}}  =  \frac{8\pi}{3}\delta({\bf r}_{{\rm
    K}}){\bm \mu}_{{\rm K}}.
\end{equation}
We have defined ${\bf r}_{{\rm K}} = {\bf R}_{{\rm K}}-{\bf r} $, where ${\bf R}_{{\rm K}}$ is the position of 
nucleus ${\rm K}$ and ${\cal I}$ is the identity matrix.
Here ${\rm V}^{(0)} ({\bf r})$ is the ground-state all-electron local potential and
 ${\rm V}^{(1)}({\bf r})$ is the corresponding first order variation. 
The latter term is due to the change in magnetisation density induced by ${\bm
  \mu}_{{\rm K}}$. The perturbation does not give rise to a first order change in the charge
 density and so there is no change in the Hartree
 potential for linear order. This can be understood by considering the effect of time
reversal; the charge density is even under time inversion, while the
spin-magnetisation and magnetic field change sign. As a result the
perturbation does not induce a first order change in the charge density,
however there is a first order induced spin-magnetisation
density. ${\rm V}^{(1)}({\bf r})$ therefore accounts for the first-order variation of the
exchange-correlation term which we label as ${\rm H}_{\rm xc}^{(1)}$.

We now use the PAW transformation
(Eqn.~\ref{eq:paw}) to obtain the pseudo-Hamiltonian. As
Eqn.~\ref{eq:paw} does not contain any field dependence it is sufficient
to apply it to each term of Eqn.~\ref{eq:ae-1st} individually. The
pseudo-Hamiltonian at zeroth order in the perturbation is
\begin{equation}\label{eq:h-zero}
\widetilde{\rm H}^{(0)} = \frac{1}{2}{\bf p}^{2} + {\rm V}_{\rm loc}({\bf r}) + 
\sum_{{\bf R}}{\rm V}^{{\bf R}}_{\rm nl},
\end{equation}
where ${\rm V}_{\rm loc}({\bf r})$ is the local part of the pseudopotentials 
which includes the self-consistent part of the Hamiltonian and 
${\rm V}^{{\bf R}}_{\rm nl}$ is the non-local part which is given by 
${\rm V}^{{\bf R}}_{\rm nl} = \sum_{n,m}\ket{\widetilde{p}_{{\bf R},n}}
a^{{\bf R}}_{n,m}\bra{\widetilde{p}_{{\bf R},n}}$. $a^{\bf R}_{n,m}$
are the strengths of the nonlocal potential in each channel at each
 ionic site. Collecting terms to linear order in the perturbation gives,
\begin{equation}\label{eq:h1_sd}
\widetilde{\rm H}^{(1)} = \widetilde{\rm H}_{\rm xc}^{(1)}+ \widetilde{\rm H}_{\rm SD} + \widetilde{\rm H}_{\rm FC}.
\end{equation}
 $\widetilde{\rm H}_{\rm SD}$ describes the spin-dipolar
 interaction induced by ${\bm \mu}_{{\rm K}}$ and is given by
\begin{equation}\label{eq:h_sd}
\widetilde{\rm H}_{\rm SD}  =   g\beta{\bf S}\cdot {\bf B}_{{\rm K}}^{\rm SD}+ 
g\beta{\bf S}\cdot \Delta {\bf B}_{{\rm K}}^{\rm SD}. 
\end{equation}  
The first term on the right-hand side is the all-electron operator and 
the second term is the augmentation to this,
\begin{eqnarray}\label{eq:delta_b_sd}
\Delta {\bf B}_{{\rm K}}^{\rm SD}& = &\sum_{n,m}\ket{\widetilde{p}_{{\bf R},n}}
\left[\right.\bracket{\phi_{{\bf R},n}}{{\bf B}_{{\rm K}}^{\rm SD}}{\phi_{{\bf
	R},m}} \\
& & - \bracket{\widetilde{\phi}_{{\bf R},n}}{{\bf B}_{{\rm K}}^{\rm SD}}
{\widetilde{\phi}_{{\bf R},m}}\left.\right]\bra{\widetilde{p}_{{\bf R},m}},   \nonumber
\end{eqnarray}
with ${\bf R} = {\bf R}_{K}$.
In Eqn.~\ref{eq:h_sd} we have only included the augmentation of the
spin-dipolar operator at the site of the perturbing atom. This on-site
approximation is fully justified given the localised nature of this
operator. 

 $\widetilde{\rm H}_{\rm FC}$ is the Fermi-contact operator and can be 
constructed in a similar manner to the spin-dipole operator giving an
all-electron and an augmentation contribution. However, as the
Fermi-contact operator contains a Dirac delta-function and is therefore
localised within the augmentation region, 
$\widetilde{\rm H}_{\rm FC}$ can be simplified considerably. 
The pseudo-partial waves
and projectors, $\sum_{n}\braket{\widetilde{p}_{n}}{\widetilde{\phi}_{n}}$,
form a complete set which enables us to rewrite the 
all-electron contribution in terms of the pseudo-partial waves within 
the augmentation region and so we can equivalently express the operator as
\begin{equation}\label{eq:h_fc}
\widetilde{\rm H}_{\rm FC}  =  g\beta{\bf S}\cdot\sum_{n,m}\ket{\widetilde{p}_{{\bf R},n}}
 \bracket{{\phi}_{{\bf R},n}}{{\bf B}_{{\rm K}}^{\rm FC}}{{\phi}_{{\bf R},m}}\bra{\widetilde{p}_{{\bf R},m}},
\end{equation}
where $\R = \R_{{\rm K}}$.
This form is more suitable for a practical calculation as it avoids an
explicit representation of the delta-function.

%%%%%%%%%%%%%%%%%%%%%%%%%%%%%%%%%%
\subsection{Magnetisation Density}
%%%%%%%%%%%%%%%%%%%%%%%%%%%%%%%%%%
To construct the magnetisation density, we define ${\bf
  m}_{i}^{(1)}({\bf r})$ to be the linear response to the magnetic
field, ${\bf B}_{i}$ induced along the direction ${\bf u}_{i}$ by the spin-dipolar and
Fermi-contact interactions.  The total magnetisation density is obtained
as ${\bf m}^{(1)} = \sum_{i=x,y,z} {\bf m}_{i}^{(1)}({\bf r})$, the
sum over the cartesian directions. By choosing ${\bf u}_{i}$ as the spin
quantisation axis, ${\rm H}^{(1)}$ is diagonal in the spin-up and
spin-down basis. The eigenstates of  ${\rm H}^{(0)} + {\rm H}^{(1)} $
are also eigenstates of ${\bf u}_{i}\cdot {\bf S}$ and so the
magnetisation density is parallel to ${\bf u}_{i}$ giving;
\begin{equation}
{\bf m}_{i}^{(1)}({\bf r}) = [{\bf u}_{i}\cdot{\bf m}_{i}^{(1)}({\bf
  r})]{\bf u}_{i} = {m}_{i}^{(1)}({\bf r}){\bf u}_{i}.
\end{equation}
Here
\begin{equation}\label{eq:m_1}
{m}_{i}^{(1)}({\bf r}) = g\beta\left[{n}_{i,\uparrow}^{(1)}({\bf r})
  - {n}_{i,\downarrow}^{(1)}({\bf r})\right] = 
2g\beta{n}_{i,\uparrow}^{(1)}({\bf r}),
\end{equation}
where g  and $\beta$ were defined previously and ${n}_{i,\sigma}^{(1)}$
is the induced density for either spin-up or spin-down.
The simplification of the magnetisation density in this way is a
consequence of time reversal symmetry, namely the absence of
 a first-order charge density. 
This means that the spin-up and spin-down ground-state wavefunctions are 
equivalent, so that $|\widetilde{\psi}^{(0)}_{\uparrow o}\rangle = 
|\widetilde{\psi}^{(0)}_{\downarrow o}\rangle$.  Also, the linear 
variation of the wavefunctions induced by the spin magnetisation are related through
$|\widetilde{\psi}^{(1)}_{\uparrow o}\rangle = -|\widetilde{\psi}^{(1)}_{\downarrow o}\rangle$. 

Within PAW, Eqn.~\ref{eq:m_1} becomes
\begin{eqnarray}\label{eq:mag_den}
{m}_{i}^{(1)}({\bf r})& = & 4g\beta {\rm Re} \sum_{ o} 
\langle\widetilde{\psi}^{(1)}_{\uparrow o}|{\bf r}
\rangle\langle{\bf r}|\widetilde{\psi}^{(0)}_{\uparrow o}\rangle \nonumber \\
 &  & + \sum_{{\bf R},n,m}
 \braket{\widetilde{\psi}_{ o\uparrow}^{(1)}}
{\widetilde{p}_{{\bf R},n}}\Bigl[
\braket{\phi_{{\bf R},n}}{{\bf r}}\braket{{\bf r}}{\phi_{{\bf R},m}} \nonumber \\ 
&  & -
\braket{\widetilde{\phi}_{{\bf R},n}}{{\bf r}}\braket{{\bf r}}{\widetilde{\phi}_{{\bf R},m}}\Bigr]
\braket{\widetilde{p}_{{\bf R},n}}{\widetilde{\psi}_{ o\uparrow}^{(0)}}. \\ \nonumber
\end{eqnarray}
${\rm Re}$ signifies taking the real component, 
$|\widetilde{\psi}^{(0)}_{\uparrow o}\rangle$ are the eigenstates of
the unperturbed Hamiltonian, ${\rm H}^{(0)}$, 
$|\widetilde{\psi}^{(1)}_{\uparrow o}\rangle$ are the perturbed
pseudowavefunctions and $ o$ indexes the occupied bands. 
The first term on the right hand side of Eqn.~\ref{eq:mag_den} is the pseudo-magnetisation
 density $\widetilde{m}_{i}^{(1)}$, and the second term is the corresponding augmentation. 
 For simplicity, we drop the spin
indexing on the ground-state wavefunctions from now on as the spin-dependence enters only 
through the perturbation.

To calculate $|\widetilde{\psi}^{(1)}_{\uparrow o}\rangle$ we employ a
Green's function method where
\begin{equation} \label{eq:green_mag}
|\widetilde{\psi}^{(1)}_{\uparrow o}\rangle = {\cal G}(\epsilon)
\ket{\psi^{(0)}_{ o}} = 
\sum_{e}\frac{\ket{\psi_{e}^{(0)}}{\bra{\psi_{
        e}^{(0)}}}}{\epsilon_{ o}- \epsilon_{e}}\widetilde{\rm
  H}_{i}^{(1)}
\ket{\psi^{(0)}_{ o}}.
\end{equation}
$\widetilde{\rm H}_{i}^{(1)}$ is the first order Hamiltonian given by
Eqn.~\ref{eq:h1_sd} with the spin quantised along the ${\bf u}_{i}$ direction.
${\cal G}(\epsilon)$ is the Green's function,
$\epsilon_{ o}$ and $\epsilon_{e}$ are the eigenvalues of the
occupied and empty bands.
 Rather 
than explicitly sum over the empty states, we project onto the 
occupied bands by multiplying 
Eqn.~\ref{eq:green_mag} through by $(\epsilon_{ o} - 
\widetilde{\rm H}^{(0)})$. We define ${\cal P} = 
\sum_{e}\ket{\widetilde{\psi}_{e}^{(0)}}\bra{\widetilde{\psi}_{e}^{(0)}} = 1 - 
\sum_{ o}\ket{\widetilde{\psi}_{ o}^{(0)}}
\bra{\widetilde{\psi}_{ o}^{(0)}}$ and rewrite Eqn.~\ref{eq:green_mag} as 
\begin{equation}\label{eq:green}
(\epsilon_{ o} - \widetilde{\rm
  H}^{(0)})\ket{\widetilde{\psi}_{\uparrow o}^{(1)}} 
= {\cal P}{\widetilde{\rm H}}_{i}^{(1)}\ket{\widetilde{\psi}_{ o}^{(0)}}.
\end{equation}
This is then solved using a conjugate gradient minimisation scheme, 
with an additional self-consistency condition to account
for the dependence of ${\rm H}_{\rm xc}^{(1)}$ on the spin-density. 
For a more detailed account of this type of approach, see Ref.~\onlinecite{gonze97}.
%%%%%%%%%%%%%%%%%%%%%%%%%%%%%%%%%%%
\subsection{Induced Magnetic Field}
%%%%%%%%%%%%%%%%%%%%%%%%%%%%%%%%%%%
The induced magnetic field at atom ${\rm L}$, and subsequently the
J-coupling between ${\rm L}$ and ${\rm K}$, due to the spin magnetisation is
obtained by combining Eqns. \ref{eq:b_ind} and \ref{eq:mag_den}
to give
\begin{equation}
{\bf B}^{(1)}_{{\bf m}^{(1)}}({\bf R}_{{\rm L}}) = \widetilde{\bf B}^{(1)}_{\rm SD}({\bf R}_{{\rm L}}) + 
\Delta{\bf B}^{(1)}_{\rm SD}({\bf R}_{{\rm L}}) + \Delta{\bf B}^{(1)}_{\rm FC}({\bf R}_{{\rm L}}),
\end{equation}
where we have taken advantage of the linearity of Eqn.~\ref{eq:b_ind} 
to yield three separate terms. The first term is the pseudo
spin-dipolar contribution and the second term is the spin-dipole 
augmentation. The notation used here implicitly assumes that a
rotation over the spin-axis has been performed.

 In practise, the pseudo-spin dipole term can be constructed from 
the Fourier transform of Eqn.~\ref{eq:b_ind}, 
\begin{equation}
{\bf B}^{(1)}_{\rm SD}({\bf G}) = -\frac{\mu_{0}}{3}\left[\frac{3(\widetilde{\bf
  m}^{(1)}({\bf G})\cdot{\bf G}){\bf G}
-\widetilde{\bf m}^{(1)}({\bf G})|{\bf G}|^{2}}{ G^{2}}\right],
\end{equation}
where $\widetilde{\bf m}^{(1)}({\bf G})$ is the Fourier transform of the 
pseudo-magnetisation density and ${\bf G}$ is a reciprocal
space lattice vector. The $G=0$ term is neglected as the cell size is
large compared with the strength of the perturbation which is small.

The induced magnetic field 
at atom ${\rm L}$ is then recovered by performing a slow inverse Fourier 
transform at the position of each responding nucleus.
The spin dipole augmentation term is obtained by using an on-site 
approximation and evaluating terms of the form,
\begin{equation}\label{eq:b_sd_aug}
 \Delta{\bf B}^{(1)}_{\rm SD}({\bf R}_{{\rm L}}) =  g\beta\frac{\mu_{0}}{2\pi}{\rm Re}
\sum_{n,m}\bracket{\widetilde{\psi}_{ o}^{(0)}}
{\Delta{\bf B}^{\rm SD}_{{\rm L}}}{\widetilde{\psi}_{\uparrow o}^{(1)}}.
\end{equation}
$\Delta{\bf B}^{\rm SD}_{{\rm L}}$ is defined in Eqn.~\ref{eq:delta_b_sd} 
but now the subscript indicates the responding 
rather than the perturbing nucleus. To evaluate this term and
Eqn.~\ref{eq:delta_b_sd}, we note 
that $\ket{\phi_{n}}$ can be decomposed into
the product of a radial ($\ket{{\rm R}_{nl}}$) and an angular 
($\ket{Y_{lm}}$) term. $B_{{\rm L}}^{\rm SD}$ can also be rewritten 
as the product of a radial and angular component such that the 
computation of the augmentation term involves the on-site calculation of
$\bracket{R_{nl}}{\frac{1}{r_{{\rm L}}^{3}}}{R_{n'l'}}\bracket{Y_{n}}
{\frac{3{\bf r}_{{\rm L}}{\bf r}_{{\rm L}}^{\rm T}}{r_{{\rm L}}^{2}} - {\cal I}}{Y_{m}}$.
The latter quantity reduces to an integral over spherical harmonics 
given by the Gaunt coefficients.

The Fermi-contact contribution, $\Delta{\bf B}^{(1)}_{\rm FC}$, is 
obtained by following the same argument used in constructing $\widetilde{\rm
  H}_{\rm FC}$ (Eqn.~\ref{eq:h_fc}) and is given by
\begin{eqnarray}
&&\Delta{\bf B}^{(1)}_{\rm FC}({\bf R}_{{\rm L}}) =  \nonumber \\
&& 
\frac{4\mu_{0}}{3}{\rm Re}\sum_{n,m}
\braket{\widetilde{\psi}_{ o}^{(0)}}{\widetilde{p}_{{\bf R},n}}
\bracket{\phi_{{\bf R},n}}{\delta({\bf
    r}_{{\rm L}})}{\phi_{{\bf R},m}}\braket{\widetilde{p}_{{\bf R},m}}
{\widetilde{\psi}_{\uparrow o}^{(1)}}, \nonumber \\  
\end{eqnarray}
which is the all-electron reconstruction of the induced magnetisation
density at the responding nucleus.

%%%%%%%%%%%%%%%%%%%%%%%%%
\section{Current Density}\label{sec:cur}
%%%%%%%%%%%%%%%%%%%%%%%%%

We now obtain the contribution to the J-coupling tensor arising from the
interaction of the nuclear spins mediated by the electron charge current. The
derivation of the current density is similar to that of the
magnetisation density and much of the notation is conserved through-out. 
We first obtain an expression for the pseudo-Hamiltonian in the presence
of a perturbing nuclear spin and show how it can be used to obtain the
induced current density. We then use this current density to calculate
the magnetic field induced at the receiving nucleus.
%%%%%%%%%%%%%%%%%%%%%%%%%%%%%%%%%%%
\subsection{Pseudo-Hamiltonian}
%%%%%%%%%%%%%%%%%%%%%%%%%%%%%%%%%%%

The all-electron Hamiltonian for a system of N magnetic nuclei 
which interact through the nuclear vector potential, ${\bf
  A}({\bf r}) = \sum_{{\rm N}}{\bf A}_{{\rm N}}({\bf r}) = 
\frac{\mu_{0}}{4\pi}\sum_{{\rm N}}\frac{{\bm \mu}_{{\rm N}}
\times{\bf r}_{{\rm N}}}{|{\bf r}_{{\rm N}}|^{3}}$, 
can be expanded to first order in ${\bm \mu}_{{\rm K}}$ to give,
\begin{equation}\label{eq:ham_orb_b}
{\rm H}_{{\rm K}} = \frac{1}{2}{\bf p}^{2} + {\rm V}^{(0)}({\bf r}) + {\rm
  H}_{{\bf A}_{\rm K}},
\end{equation}
where 
\begin{equation}
{\rm H}_{{\bf A}_{{\rm K}}} = \frac{\mu_{0}}{4\pi}{\bf p}\cdot{\bf A}_{{\rm K}}({\bf r}),
\end{equation}
and ${\rm V}^{(0)}({\bf r})$ is the ground-state local potential. The perturbation does not 
induce a first order change in either the charge or magnetisation
densities and so unlike Eqn.~\ref{eq:ae-1st}, there
is no linear variation to the self-consistent potential in Eqn.~\ref{eq:ham_orb_b}.
We have used the symmetric gauge for the vector potential and taken the natural choice of
gauge-origin; namely that for the ${\rm N}$th nuclear spin the gauge origin is
the ${\rm N}$th atomic site giving ${\bf A}_{{\rm N}}({\bf
  r})=1/2 {\bf B}({\bf r}) \times {\bf r}_{\rm N}$.
This gauge-choice preserves the translational invariance of the system 
and is much simpler than in 
the otherwise analogous case of NMR chemical shielding. 
In the latter situation, due to the use of finite basis sets
a rigid translation of the system in the 
uniform external magnetic field introduces an additional phase factor.
For the planewave-pseudopotential approach the problem was solved by 
Pickard and Mauri with the development of the Gauge-Including Projector 
Augmented-Wave (GIPAW) approach, an extension which is unnecessary here.

To obtain the pseudo-Hamiltonian we now apply the PAW transformation of
Eqn.~\ref{eq:paw} to Eqn.~\ref{eq:ham_orb_b}. The zeroth order term is again
given by Eqn.~\ref{eq:h-zero} and the first order term by
\begin{eqnarray}\label{eq:orb_ham}
\widetilde{\rm H}_{{\bf A}_{{\rm K}}}& = &
\frac{\mu_{0}}{4\pi} \bm{\mu}_{\rm K}\cdot 
\frac{{\bf r}_{{\rm K}}\times{\bf p}}{|{\bf r}_{{\rm K}}|^{3}} + \nonumber \\
& &\frac{\mu_{0}}{4\pi} \bm{\mu}_{\rm K}\cdot
\sum_{n,m}\ket{{\widetilde p}_{{\bf R},n}}
 \left[\bracket{\phi_{{\bf R},n}}{\frac{{\bf L}_{{\rm K}}}{|{\bf r}_{{\rm K}}|^{3}}}
{\phi_{{\bf R},m}}\right. \nonumber \\ 
& & -  \left.\bracket{\widetilde{\phi}_{{\bf R},n}}
{\frac{{\bf L}_{{\rm K}}}{|{\bf r}_{{\rm K}}|^{3}}}{\widetilde{\phi}_{{\bf R},m}}\right]
\bra{\widetilde{p}_{{\bf R},m}}.
\end{eqnarray}
The first term on the right-hand side is the all-electron component 
and the second term is the augmentation which is constructed at
the site of the perturbing nucleus. 
${\bf L}_{\rm K} = {\bf r}_{{\rm K}}\times {\bf p}$ is the angular momentum 
operator centred on the perturbing atomic site.

%%%%%%%%%%%%%%%%%%%%%%%%%%%%%%%%
\subsection{Current Density}
%%%%%%%%%%%%%%%%%%%%%%%%%%%%%%%%
The current density operator, ${\bf J}({\bf r})$, is given by the 
sum of a paramagnetic and a diamagnetic term, 
\begin{equation}
{\bf J}({\bf r}) = {\bf J}^{\rm p}({\bf r}) + {\bf J}^{\rm d}({\bf r}),
\end{equation}
where the paramagnetic term is given by
\begin{equation}\label{eq:jp_op}
{\bf J}^{\rm p}({\bf r}) = -\left[{\bf p}\ket{\bf r}\bra{\bf r} + 
\ket{\bf r}\bra{\bf r}{\bf p}\right]/2,
\end{equation}
and the diamagnetic term is 
\begin{equation}
{\bf J}^{\rm d}({\bf r}) = -{\bf A}({\bf r})\ket{\bf r}\bra{\bf r}.
\end{equation}
If we consider the current due only to the perturbing nucleus, and with
our atomic choice of gauge origin, the diamagnetic
term can be written as
\begin{equation}\label{eq:jd_op}
{\bf J}^{\rm d}_{{\rm K}}({\bf r}) = -\frac{\mu_{0}}{4\pi}\frac{{\bm \mu}_{{\rm K}}\times{\bf r}_{{\rm K}}}{r_{{\rm K}}^{3}}
\ket{\bf r}\bra{\bf r}.
\end{equation}
By applying Eqn.~\ref{eq:paw} to both Eqns. \ref{eq:jp_op} and \ref{eq:jd_op}, 
we obtain the pseudo-current density operator within PAW
\begin{equation}
\widetilde{\bf J}({\bf r}) = {\bf J}^{\rm p}({\bf r}) + {\bf J}^{\rm
  d}_{{\rm K}}({\bf r}) 
+ \left[\Delta{\bf J}^{\rm p}({\bf r}) + 
\Delta{\bf J}^{\rm d}_{{\rm K}}({\bf r})\right],
\end{equation}
where the paramagnetic augmentation operator is
\begin{eqnarray}
\Delta{\bf J}^{\rm p}({\bf r})& = &\sum_{{\bf R},n,m}\ket{\widetilde{p}_{{\bf
      R},n}}\left[\bracket{\phi_{{\bf R},n}}{{\bf J}^{\rm p}}
{\phi_{{\bf R},m}}\right. \nonumber \\ 
&& -\bracket{\widetilde{\phi}_{{\bf R},n}}{{\bf J}^{\rm p}}
{\widetilde{\phi}_{{\bf R},m}}\left.\right]\bra{\widetilde{p}_{{\bf R},m}},
\end{eqnarray}
and the corresponding diamagnetic operator is
\begin{eqnarray}
\Delta{\bf J}^{\rm d}_{{\rm K}}({\bf r})& = & \sum_{{\bf R},n,m}\ket{\widetilde{p}_{{\bf
      R},n}}
\left[\bracket{\phi_{{\bf R},n}}{{\bf J}^{\rm d}_{{\rm K}} }{\phi_{{\bf R},m}}\right. \nonumber \\
& & -\bracket{\widetilde{\phi}_{{\bf R},n}}{{\bf J}^{\rm d}_{{\rm K}}}
{\widetilde{\phi}_{{\bf R},m}}\left.\right] \bra{\widetilde{p}_{{\bf R},m}}.
\end{eqnarray}
Arranging terms in ${\bf J}({\bf r})$ to zeroth and linear order in
${\bm \mu}_{\rm K}$ gives
\begin{equation}\label{eq:j_zero}
{\bf J}^{(0)}({\bf r}) = {\bf J}^{\rm p}({\bf r}) + 
\Delta{\bf J}^{\rm p}({\bf r}),
\end{equation}
and
\begin{equation}\label{eq:j_one}
{\bf J}^{(1)}({\bf r}) =  {\bf J}^{\rm d}_{{\rm K}}({\bf r}) + 
\Delta{\bf J}^{\rm d}_{{\rm K}}({\bf r}).
\end{equation}
Using Eqns.~\ref{eq:j_zero} and \ref{eq:j_one} we are now able to obtain
the first-order induced current density, which within density functional
perturbation theory is given by
\begin{eqnarray}\label{eq:current}
{\bf j}^{(1)}({\bf r})& = & 2\sum_{o} 2\,{\rm Re}
\bracket{\widetilde{\psi}_{o}^{(0)}}{\widetilde{{\bf J}}^{(0)}}
{\widetilde{\psi}_{o}^{(1)}} 
+ \bracket{\widetilde{\psi}_{o}^{(0)}}{\widetilde{\bf J}^{(1)}}
{\widetilde{\psi}_{o}^{(0)}}, \nonumber \\
& = & {\bf j}^{(1)}_{\rm p}({\bf r}) + {\bf j}^{(1)}_{\rm d}({\bf r}).
\end{eqnarray}
Here $\ket{\widetilde{\psi}_{o}^{(0)}}$ is the unperturbed wavefunction,
$\ket{\widetilde{\psi}_{o}^{(1)}}$ is the perturbed wavefunction and
$o$ indexes the occupied bands. The first term on the right 
hand side is the paramagnetic contribution to the induced current and
the second term is the diamagnetic contribution. 
The first order wavefunction is again obtained using Eqn.~\ref{eq:green}
where $\widetilde{\rm H}^{(1)}$ is now given by Eqn.~\ref{eq:orb_ham}.
%%%%%%%%%%%%%%%%%%%%%%%%%%%%%%%%%%%%%%%
\subsection{Induced Magnetic Field}
%%%%%%%%%%%%%%%%%%%%%%%%%%%%%%%%%%%%%%%
%
We can now combine Eqns.~\ref{eq:b_ind} and \ref{eq:current} to 
calculate the J-coupling between nuclei ${\rm L}$ and 
${\rm K}$ arising from the magnetic field induced by the orbital 
current. The magnetic field at nucleus ${\rm L}$ due to the orbital
current can be expressed as the sum of 4 terms,
\begin{eqnarray}\label{eq:b_orb}
{\bf B}^{(1)}_{{\bf j}^{(1)}}({\bf R}_{{\rm L}})& = &\widetilde{\bf B}_{\rm
  p}^{(1)}({\bf R}_{{\rm L}}) + \widetilde{\bf B}_{\rm d}^{(1)}({\bf R}_{{\rm L}}) +
\Delta{\bf B}_{\rm p}^{(1)}({\bf R}_{{\rm L}}) \nonumber \\
 && + \Delta{\bf B}_{\rm d}^{(1)}({\bf R}_{{\rm L}}), 
\end{eqnarray}
the pseudised contributions from 
the paramagnetic and diamagnetic currents and their 
respective augmentation terms. 

To calculate the pseudised contributions to the current density we 
obtain the Fourier transform of 
$\widetilde{\bf B}_{\rm p}^{(1)}({\bf R}_{{\rm L}})$ and  
$\widetilde{\bf B}_{\rm d}^{(1)}({\bf R}_{{\rm L}})$ giving
\begin{equation}\label{eq:cur-for}
{\bf B}^{(1)}({\bf G}) = \mu_{0}\frac{i{\bf G}\times
{\bf j}_{\rm p/d}^{(1)}({\bf G})}{G^{2}},
\end{equation}
where ${\bf j}_{\rm p/d}^{(1)}({\bf G})$ is the Fourier transform of 
either the paramagnetic or diamagnetic current. To obtain 
the induced field at the atom site ${\rm L}$ we perform a slow Fourier transform of
Eqn.~\ref{eq:cur-for}. We again note that the G=0 contribution to
${\bf B}^{(1)}({\bf G})$ is neglected as the contribution is expected
to be small. The augmentation to the paramagnetic current is calculated using 
an on-site approximation (${\bf R} = {\bf R}_{\rm L}$) with
\begin{eqnarray}
\Delta{\bf B}_{\rm p}^{(1)}({\bf R}_{{\rm L}})& = & \frac{\mu_{0}}{4\pi}\sum_{n,m}
\braket{\widetilde{\psi}_{o}^{(0)}}{\widetilde{p}_{{\bf R},n}}
\left[\bracket{\phi_{{\bf R},n}}{\frac{{\bf L}_{\rm L}}{{\bf r}_{{\rm L}}^{3}}}
{\phi_{{\bf R},m}}\right.\nonumber \\ 
&& -\left.\bracket{\widetilde{\phi}_{{\bf R},n}}{\frac{{\bf L}_{\rm L}}{{\bf r}_{{\rm L}}^{3}}}
{\widetilde{\phi}_{{\bf R},m}}\right]
\braket{\widetilde{p}_{{\bf R},m}}{\widetilde{\psi}_{o}^{(1)}},
\end{eqnarray}
where ${\bf L}$ is the angular momentum operator evaluated with 
respect to the augmentation regions. The augmentation to the 
diamagnetic current is given by
\begin{eqnarray}
\Delta{\bf B}_{\rm d}^{(1)}& = &  \frac{\mu_{0}}{4\pi}\sum_{{\bf
    R},n,m}
\braket{\widetilde{\psi}_{o}^{(0)}}{\widetilde{p}_{{\bf R},n}} \nonumber \\
&& \left[\bracket{\phi_{{\bf R},n}}{\frac{ r_{{\rm L}} r_{{\rm K}} - 
{\bf r}_{{\rm L}}{\bf r}_{{\rm K}}^{\rm T}}{r_{{\rm L}}^{3}r_{{\rm K}}^{3}}}{\phi_{m}}\right.\nonumber \\
&-&  \left.\bracket{\widetilde{\phi}_{{\bf R},n}}{\frac{ r_{{\rm L}} r_{{\rm K}} - 
{\bf r}_{{\rm L}}{\bf r}_{{\rm K}}^{\rm T}}{r_{{\rm L}}^{3}r_{{\rm K}}^{3}}}{\widetilde{\phi}_{m}}\right]
\braket{\widetilde{p}_{{\bf R},m}}{\widetilde{\psi}_{o}^{(1)}}.\nonumber \\
\end{eqnarray}
This is much more difficult to evaluate than any other term as an
on-site approximation cannot be used due to the presence
of the ${\bf r}_{{\rm K}}$ position vector within the augmentation
summation.
However, previous quantum chemical studies have shown that 
the overall contribution to the J-coupling 
from the diamagnetic term is very small compared with any of the 
other three contributions. In light of this, we have neglected this 
term in our current implementation.

%%%%%%%%%%%%%%%%%%
\section{Results}\label{sec:res}
%%%%%%%%%%%%%%%%%%

We have implemented our theory into a parallelised plane-wave 
electronic structure code.\cite{clark05} The ground-state wavefunctions and 
Hamiltonian are obtained self-consistently after which the isotropic
J-coupling constant is calculated using the outlined approach. 
In our implementation we use norm-conserving Troullier-Martins 
pseudopotentials.\cite{troullier1} For the all-electron reconstruction we used two 
projectors per angular momentum channel. 

In the following sections we compare our approach to
existing quantum chemistry approaches which use localised basis sets 
and with experiment. 

%%%%%%%%%%%%%%%%%%%%%%%
\subsection{Molecules}
%%%%%%%%%%%%%%%%%%%%%%%

To validate our method we have calculated isotropic coupling constants for
a range of small molecules\cite{notemol} and compared them with experiment. 
There are several studies of calculated J-couplings for small molecules reported in 
the literature, using a variety of theoretical approaches, see
Ref.~\onlinecite{vaara02} and references therein. We compare our results to
calculations presented by Lantto {\it et al}\cite{lantto02} which were
obtained within DFT using the BLYP functional and with the
Multi-configurational Self-Consistent Field (MCSCF) approach.
For consistency we use the molecular geometries reported in 
their work.

To obtain the isotropic J-coupling we use a supercell of size 1728 \AA$^{3}$ for each 
molecule with the exception of benzene which required a larger cell-size of 
3375 \AA$^{3}$. The exchange-correlation was approximated by GGA-PBE\cite{perdew1} and an energy cut-off
of 80 Ry was imposed on the planewave expansion. All calculations
sample the Brillouin zone at the gamma-point and used norm-conserving 
Trouillier-Martins pseudopotentials.\cite{troullier1}
\begin{figure}
\includegraphics*[width=8.0cm]{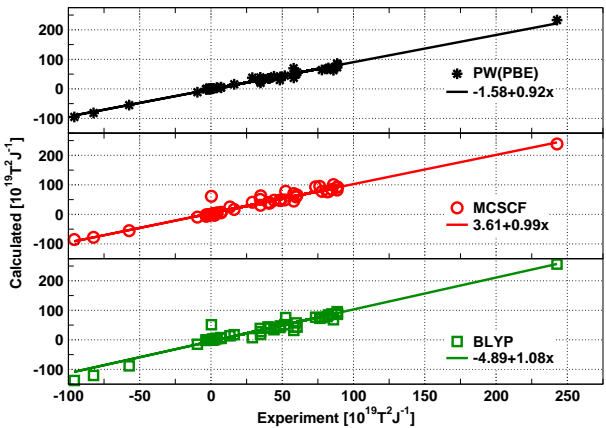}
\caption{\label{fig:molecules}J-couplings calculated for a set of molecules. Both the MCSCF 
and BLYP results were taken from Ref.~\onlinecite{lantto02}. All of the
experimental values were also taken from this paper. The PW(PBE) results are from the 
present work. The lines are obtained from a linear 
regression of the calculated values with experiment.  
All values are quoted in $10^{19}{\rm T}^{2} {\rm J}^{-1}$, the unit of the 
reduced coupling constant.}
\end{figure}
The calculated J-coupling against experiment for the molecules are shown in Fig. 
\ref{fig:molecules} alongside the results of a linear regression for each set of data. 
These results are presented as reduced spin-coupling constants, which are given by 
$\overleftrightarrow{\bf K}_{{\rm LK}} = \frac{2\pi\overleftrightarrow{\bf J}_{{\rm LK}}}{\hbar
\gamma_{{\rm L}}\gamma_{{\rm K}}}$, and so are independent of nuclear species. 
The graph indicates an excellent overall agreement with both experiment and 
the other approaches. The accuracy of the planewave PBE calculations is comparable
with the all-electron BLYP results, with regression coefficients of
0.92 and 1.08 respectively. The correlation coefficients for the BLYP, MCSCF
and PW data are 0.97, 0.97 and 0.99 respectively, suggesting a smaller random
error in the planewave approach. Unsurprisingly, the most accurate couplings are 
given by MCSCF, which provides a more comprehensive description of electron 
correlation but is computationally more demanding than DFT. 

\begin{table}
 \caption{\label{tab:f_ben}J-coupling[Hz] in benzene. PW(PBE) labels the current 
planewave approach. The BLYP, MCSCF and experimental values were taken from Ref.~\onlinecite{lantto02}.
D, P, FC and SD label the diamagnetic, paramagnetic, Fermi-contact and spin-dipolar contributions respectively. All values are in Hz.}
\begin{ruledtabular} 
\begin{tabular}{lcdddddd}
%\begin{tabular}{lccccccc} 
$J^{n}_{{\rm KL}}$&
Method&
\multicolumn{1}{c}{$\mathrm{D}$}&
\multicolumn{1}{c}{$\mathrm{P}$}&
\multicolumn{1}{c}{$\mathrm{FC}$}&
\multicolumn{1}{c}{$\mathrm{SD}$}&
\multicolumn{1}{c}{$\mathrm{Total}$}&
\multicolumn{1}{c}{$\mathrm{Expt[Hz]}$} \\ \hline
J$_{\rm CC}^{1}$   &PW(PBE)& 0.2   &-6.9   & 58.2  & 1.9  & 53.4  &55.8     \\  
		  &BLYP   & 0.2  &-6.8   & 63.8  & 1.0  & 58.2    &       \\
		  &MCSCF  & 0.4  &-6.6   & 75.1  & 1.5  & 70.9    &       \\
J$_{\rm CC}^{2}$    &PW(PBE)& 0.0 & 0.0   & -1.2   & -0.3 & -1.5  &-2.5     \\  
		  &BLYP   & 0.0  & 0.1   & 0.0   & -0.5 & -0.4    &       \\
		  &MCSCF  &-0.2  & 0.0   & -3.7  & -1.1 & -5.0    &       \\
J$_{\rm CC}^{3}$    &PW(PBE)& 0.0 & 0.6   & 7.3    & 1.1  & 9.0   &10.1     \\  
		  &BLYP   &0.0   & 0.5   & 8.4   & 1.5  & 10.4    &       \\
		  &MCSCF  &0.1   & 0.4   & 16.8  & 1.8  & 19.1    &       \\
J$_{\rm CH}^{1}$   &PW(PBE)& 0.5  & 0.9   &132.5  &-0.2  & 133.7  & 158.3  \\  
		  &BLYP   &     &       &       &      & 155.2    &       \\
		  &MCSCF  &     &       &       &      & 185.1    &       \\
J$_{\rm CH}^{2}$   &PW(PBE)&-0.1  & 0.2   & 5.1   & 0.1  & 5.3    & 1.0    \\  
		  &BLYP   &     &        &       &      & 1.1   &       \\
		  &MCSCF  &     &        &       &      & -9.8  &       \\
J$_{\rm CH}^{3}$   &PW(PBE)&-0.2  & -1.1  & 6.0   & 0.0  & 4.7    & 7.6    \\  
		  &BLYP   &     &       &       &       & 7.4    &       \\
		  &MCSCF  &     &       &       &       & 12.9    &       \\
J$_{\rm CH}^{4}$   &PW(PBE)&-0.1  & 0.3   & -0.4  & 0.0  & -0.2   & -1.2   \\  
		  &BLYP   &     &       &       &       & -1.3   &       \\
		  &MCSCF  &     &       &       &       & -6.1   &       \\
\end{tabular}
\end{ruledtabular}
\end{table}							          

In Table \ref{tab:f_ben} we present the J-coupling values calculated for benzene. 
The results compare favourably with both the existing approaches and with 
experiment. The MCSCF approach systematically overestimates the J-coupling 
for both J$_{\rm CH}$ and J$_{\rm CC}$ compared with experiment. This is due to
the use of a restricted basis set which was necessary given the size of the system,
for further details see Ref.~\onlinecite{kaski96}. 

 The decomposition of the J-coupling into the four components 
serves as an illustration of the relative strengths of each contribution and the 
trends over several bonds. Lantto {\it et al} have only presented this separation for 
J$_{\rm CC}$. It is clear that the Fermi-contact is the dominant mechanism
in the coupling and that the diamagnetic component is consistently the smallest 
and is often negligible. 

%%%%%%%%%%%%%%%%%%%%%
\subsection{Crystals}
%%%%%%%%%%%%%%%%%%%%%

Due to the difficulties encountered measuring J-coupling in solid-state systems there
are very few values to be found in the literature that are suitable for validation of our
approach. Recently Coelho {\it et al}.\cite{coelho06} provided an
estimate for the two bond coupling between $^{29}$Si and $^{31}$P pairs
in the silicophosphate Si$_{5}$O(PO$_{4}$)$_{6}$. Subsequently this was
followed by a more accurate determination \cite{coelho07} which
identified the four Si-O-P couplings. We have
calculated NMR chemical shifts and $J^{2}_{\rm P-O-Si}$ for
Si$_{5}$O(PO$_{4}$)$_{6}$ to validate our approach. 

The structure of Si$_{5}$O(PO$_{4}$)$_{6}$ is trigonal (a=7.869\AA,
c=24.138\AA, 36 atoms per primitive cell) and contains one unique 
P site and three inequivalent Si sites. Two of these Si sites are 
6-fold coordinated, Si$_{1}$ and Si$_{2}$, and the third site, 
Si$_{3}$, is four-fold coordinated. Si$_{1}$ is bonded to six
equivalent oxygen atoms, Si$_{2}$ is bonded to six oxygen atoms 
which are comprised of two distinct sites. Si$_{3}$ is bonded to three 
equivalent oxygen atoms and one oxygen from an Si$_{3}$-O
tetrahedron. Thus there is one $^{31}$P chemical shift, 
three $^{29}$Si chemical shifts and four unique $^{2}$J$_{\rm P-O-Si}$ couplings.

We obtained the structure from the Chemical Database Service at Daresbury.\cite{cds} 
Prior to calculating the NMR parameters, we performed a full geometry 
optimisation on the structure, using a planewave cut-off of 70 Ryd and norm-conserving pseudopotentials.
The GGA-PBE\cite{perdew1} exchange-correlation functional was used and
a Monkhorst-Pack k-point grid with a maximum of 0.1 \AA$^{-1}$ between sampling points. 
We calculated the NMR chemical shifts using the GIPAW\cite{pickard01} approach 
with the same parameters used for the geometry optimisation. 
The J-coupling between P and Si was obtained using the approach
outlined above. A slightly higher maximum planewave energy (80Ryd) was
required to give J-couplings converged to within 0.1Hz.

We tested the convergence of the induced magnetization density and
current density with respect to supercell size. The results of
these calculations for three cell sizes are presented in table
\ref{tab:sipo_cell}. From this we can see that both the induced
magnetization and current densities have decayed substantially within
the single unit cell. The largest of these calculations (144 atoms) was
parallelised over 16 dual-core AMD processors and took 45 hours to
run. The groundstate calculation took approximately 14 hours and the 
J-coupling terms; Fermi-contact, spin-dipolar and orbital, required
3.5 hours, 15.5 hours and 11.4 hours respectively. 

\begin{table}
 \caption{\label{tab:sipo_cell}Calculated J-coupling for silicophosphate 
Si$_{5}$O(PO$_{4}$)$_{6}$ using the unit cell and two supercells
 constructed with 2$\times$1$\times$1 and 2$\times$2$\times$1 unit cells.}
\begin{ruledtabular} 
\begin{tabular}{llll} 
Coupling                    & 1$\times$1$\times$1 & 2$\times$1$\times$1 & 2$\times$2$\times$1   \\\hline\hline
$^{2}$J$_{\rm P-O_{3}-Si_{1}}$ & -17.37 & -17.07 & -17.12\\
$^{2}$J$_{\rm P-O_{2}-Si_{2}}$ & -16.16 & -16.18 & -16.26\\
$^{2}$J$_{\rm P-O_{5}-Si_{2}}$ & -1.30  & -1.20  & -1.17 \\
$^{2}$J$_{\rm P-O_{4}-Si_{3}}$ & -13.83 & -14.18 & -14.13\\
\end{tabular}
\end{ruledtabular}
\end{table}							

The results for the 2$\times$2$\times$1 cells are presented in comparison with
experiment in Table \ref{tab:sipo}.
\begin{table}
 \caption{\label{tab:sipo}Calculated NMR chemical
   shifts\cite{noteref} and J-coupling for silicophosphate 
Si$_{5}$O(PO$_{4}$)$_{6}$. The experimental values are in brackets and
were taken from Ref.~\onlinecite{coelho07}.}
\begin{ruledtabular} 
\begin{tabular}{llll} 
Coupling                   & $^{31}$P [ppm] & $^{29}$Si [ppm]  & Calc.  [Hz] \\ \hline
${J}^{2}_{\rm P-O_{3}-Si_{1}}$& -47.4 (-43.8) &  -214.8 (-213.3)  & -17.12 (15$\pm$2)\\  
${J}^{2}_{\rm P-O_{2}-Si_{2}}$&               &  -218.7 (-217.0)  & -16.26 (14$\pm$2)\\
${J}^{2}_{\rm P-O_{5}-Si_{2}}$&               &  -218.7 (-217.0)  & -1.17  (4$\pm2$) \\
${J}^{2}_{\rm P-O_{4}-Si_{3}}$&               &  -128.6 (-119.1)  & -14.13 (12$\pm2$)\\
\end{tabular}
\end{ruledtabular}
\end{table}							
From Table \ref{tab:sipo} it is clear that the calculated J-couplings
are in excellent agreement with experiment and fully reproduce the
surprisingly large spread in the J-coupling values. Our calculations
verify the novel experimental work and also identify the sign of the
couplings which are not determined by the experimental spin-echo based
approaches.

The NMR chemical shifts are also in good agreement
with experiment, particularly for $^{29}$Si. For both $^{29}$Si and $^{31}$P the difference 
between the calculated and experimental values is a very small fraction
of the total shift range. We note that our assignment of the three Si
sites in  Si$_{5}$O(PO$_{4}$)$_{6}$ agrees with the assignment based on
experimental intensities as discussed by Coelho {\it et
  al}\cite{coelho06}. 
\begin{table}
 \caption{\label{tab:sipo2}Decomposition for the J-coupling in
   Si$_{5}$O(PO$_{4}$)$_{6}$. D is the diamagnetic term, P is the
paramagnetic term, SD is the spin-dipolar and FC is the Fermi-contact.}
\begin{ruledtabular} 
\begin{tabular}{lccccc} 
Coupling                      & D    &  P    & SD    &  FC   & Total \\ \hline
$^{2}$J$_{\rm P-O_{3}-Si_{1}}$   &-0.05 &-0.27  &-0.03 & -16.77 & -17.12\\  
$^{2}$J$_{\rm P-O_{2}-Si_{2}}$   &-0.02 & -0.50 & -0.23 &-15.51  & -16.26\\
$^{2}$J$_{\rm P-O_{5}-Si_{2}}$   &-0.10 & -0.07 & 0.18 &-1.18  & -1.17\\
$^{2}$J$_{\rm P-O_{4}-Si_{3}}$   &-0.09 & -0.49 & 0.23 &-13.79 & -14.13\\   
\end{tabular}
\end{ruledtabular}
\end{table}							
In Table~\ref{tab:sipo2} we present the decomposition of the silicophosphate J-coupling into their
constituent terms. As with benzene, the Fermi-contact is found to be consistently the largest 
component while the diamagnetic and spin-dipolar contributions are
very small.

%%%%%%%%%%%%%%%%%%%%%
\section{Conclusions}\label{sec:conc}
%%%%%%%%%%%%%%%%%%%%%
We have developed an all-electron approach for calculating NMR 
J-coupling constants using planewaves and pseudopotentials within
DFT. Our method is applicable to both solution and solid state systems 
 using supercell techniques. We have validated our theory against
existing quantum chemical approaches and experiment for molecules. We
have calculated the J-coupling between Si and P in a silicophosphate polymorph,
for which we have determined the sign of the coupling.

Given the recent experimental interest in J-coupling, we expect that 
our approach will prove useful in determining both the range and 
strength of coupling in systems not yet investigated and whether or 
not such couplings can feasibly be determined by experiment. By combining J-coupling
calculations with computations of other NMR
parameters, there now exists a comprehensive set of computational tools
to complement experimental understanding and design. 

%%%%%%%%%%%%%%%%%%%%%
\section{Acknowledgments}
%%%%%%%%%%%%%%%%%%%%%
SAJ would like to acknowledge postdoctoral funding by TCM Group under 
Grant No. S61263/01 and Science Foundation Ireland.
JRY thanks Corpus Christi College for a research fellowship.
Computational facilities were provided by the Tyndall National
Institute and the SFI/HEA Irish Centre for High-End 
Computing (ICHEC).
We would also like to thank S.P. Brown for valuable discussions 
on J-coupling experiments.


\begin{thebibliography}{48}
\expandafter\ifx\csname natexlab\endcsname\relax\def\natexlab#1{#1}\fi
\expandafter\ifx\csname bibnamefont\endcsname\relax
  \def\bibnamefont#1{#1}\fi
\expandafter\ifx\csname bibfnamefont\endcsname\relax
  \def\bibfnamefont#1{#1}\fi
\expandafter\ifx\csname citenamefont\endcsname\relax
  \def\citenamefont#1{#1}\fi
\expandafter\ifx\csname url\endcsname\relax
  \def\url#1{\texttt{#1}}\fi
\expandafter\ifx\csname urlprefix\endcsname\relax\def\urlprefix{URL }\fi
\providecommand{\bibinfo}[2]{#2}
\providecommand{\eprint}[2][]{\url{#2}}

\bibitem[{\citenamefont{Levitt}(2001)}]{levitt}
\bibinfo{author}{\bibfnamefont{M.~H.} \bibnamefont{Levitt}},
  \emph{\bibinfo{title}{Spin Dynamics. Basics of Nuclear Magnetic Resonance}}
  (\bibinfo{publisher}{Wiley}, \bibinfo{year}{2001}).

\bibitem[{\citenamefont{Duma et~al.}(2004)\citenamefont{Duma, Lai, Carravetta,
  Emsley, Brown, and Levitt}}]{duma04}
\bibinfo{author}{\bibfnamefont{L.}~\bibnamefont{Duma}},
  \bibinfo{author}{\bibfnamefont{W.~C.} \bibnamefont{Lai}},
  \bibinfo{author}{\bibfnamefont{M.}~\bibnamefont{Carravetta}},
  \bibinfo{author}{\bibfnamefont{L.}~\bibnamefont{Emsley}},
  \bibinfo{author}{\bibfnamefont{S.~P.} \bibnamefont{Brown}}, \bibnamefont{and}
  \bibinfo{author}{\bibfnamefont{M.~H.} \bibnamefont{Levitt}},
  \bibinfo{journal}{ChemPhysChem} \textbf{\bibinfo{volume}{5}},
  \bibinfo{pages}{815} (\bibinfo{year}{2004}).

\bibitem[{\citenamefont{Amoureux et~al.}(2005)\citenamefont{Amoureux, Trebosc,
  Wiench, Massiot, and Pruski}}]{amoureux05}
\bibinfo{author}{\bibfnamefont{J.~P.} \bibnamefont{Amoureux}},
  \bibinfo{author}{\bibfnamefont{J.}~\bibnamefont{Trebosc}},
  \bibinfo{author}{\bibfnamefont{J.~W.} \bibnamefont{Wiench}},
  \bibinfo{author}{\bibfnamefont{D.}~\bibnamefont{Massiot}}, \bibnamefont{and}
  \bibinfo{author}{\bibfnamefont{M.}~\bibnamefont{Pruski}},
  \bibinfo{journal}{Solid State Nucl. Magn. Reson.}
  \textbf{\bibinfo{volume}{27}}, \bibinfo{pages}{228} (\bibinfo{year}{2005}).

\bibitem[{\citenamefont{Coelho et~al.}(2006)\citenamefont{Coelho, Azais,
  Bonhomme-Coury, Maquet, Massiot, and Bonhomme}}]{coelho06}
\bibinfo{author}{\bibfnamefont{C.}~\bibnamefont{Coelho}},
  \bibinfo{author}{\bibfnamefont{T.}~\bibnamefont{Azais}},
  \bibinfo{author}{\bibfnamefont{L.}~\bibnamefont{Bonhomme-Coury}},
  \bibinfo{author}{\bibfnamefont{J.}~\bibnamefont{Maquet}},
  \bibinfo{author}{\bibfnamefont{D.}~\bibnamefont{Massiot}}, \bibnamefont{and}
  \bibinfo{author}{\bibfnamefont{C.}~\bibnamefont{Bonhomme}},
  \bibinfo{journal}{J. Mag. Res.} \textbf{\bibinfo{volume}{179}},
  \bibinfo{pages}{114} (\bibinfo{year}{2006}).

\bibitem[{\citenamefont{Cadars et~al.}(2007)\citenamefont{Cadars, Lesage,
  Trierweiler, Heux, and Emsley}}]{cadars07}
\bibinfo{author}{\bibfnamefont{S.}~\bibnamefont{Cadars}},
  \bibinfo{author}{\bibfnamefont{A.}~\bibnamefont{Lesage}},
  \bibinfo{author}{\bibfnamefont{M.}~\bibnamefont{Trierweiler}},
  \bibinfo{author}{\bibfnamefont{L.}~\bibnamefont{Heux}}, \bibnamefont{and}
  \bibinfo{author}{\bibfnamefont{L.}~\bibnamefont{Emsley}},
  \bibinfo{journal}{Phys. Chem. Chem. Phys.} \textbf{\bibinfo{volume}{9}},
  \bibinfo{pages}{92} (\bibinfo{year}{2007}).

\bibitem[{\citenamefont{Coelho et~al.}(2007)\citenamefont{Coelho, Azais,
  Bonhomme-Coury, Laurent, and Bonhomme}}]{coelho07}
\bibinfo{author}{\bibfnamefont{C.}~\bibnamefont{Coelho}},
  \bibinfo{author}{\bibfnamefont{T.}~\bibnamefont{Azais}},
  \bibinfo{author}{\bibfnamefont{L.}~\bibnamefont{Bonhomme-Coury}},
  \bibinfo{author}{\bibfnamefont{G.}~\bibnamefont{Laurent}}, \bibnamefont{and}
  \bibinfo{author}{\bibfnamefont{C.}~\bibnamefont{Bonhomme}},
  \bibinfo{journal}{Inorg. Chem.} \textbf{\bibinfo{volume}{46}},
  \bibinfo{pages}{1379} (\bibinfo{year}{2007}).

\bibitem[{\citenamefont{Brown et~al.}(2002{\natexlab{a}})\citenamefont{Brown,
  Perez-Torralba, Sanz, Claramunt, and Emsley}}]{brown02}
\bibinfo{author}{\bibfnamefont{S.~P.} \bibnamefont{Brown}},
  \bibinfo{author}{\bibfnamefont{M.}~\bibnamefont{Perez-Torralba}},
  \bibinfo{author}{\bibfnamefont{D.}~\bibnamefont{Sanz}},
  \bibinfo{author}{\bibfnamefont{R.~M.} \bibnamefont{Claramunt}},
  \bibnamefont{and} \bibinfo{author}{\bibfnamefont{L.}~\bibnamefont{Emsley}},
  \bibinfo{journal}{Chem. Commun.} pp. \bibinfo{pages}{1852--1853}
  (\bibinfo{year}{2002}{\natexlab{a}}).

\bibitem[{\citenamefont{Brown et~al.}(2002{\natexlab{b}})\citenamefont{Brown,
  Perez-Torralba, Sanz, Claramunt, and Emsley}}]{brown02b}
\bibinfo{author}{\bibfnamefont{S.~P.} \bibnamefont{Brown}},
  \bibinfo{author}{\bibfnamefont{M.}~\bibnamefont{Perez-Torralba}},
  \bibinfo{author}{\bibfnamefont{D.}~\bibnamefont{Sanz}},
  \bibinfo{author}{\bibfnamefont{R.~M.} \bibnamefont{Claramunt}},
  \bibnamefont{and} \bibinfo{author}{\bibfnamefont{L.}~\bibnamefont{Emsley}},
  \bibinfo{journal}{J. Am. Chem. Soc.} \textbf{\bibinfo{volume}{124}},
  \bibinfo{pages}{1152} (\bibinfo{year}{2002}{\natexlab{b}}).

\bibitem[{\citenamefont{Lai et~al.}(2006)\citenamefont{Lai, McLean, Gansmuller,
  Verhoeven, Antonioli, Carravetta, Duma, Bovee-Geurts, Johannessen, de~Groot
  et~al.}}]{lai06}
\bibinfo{author}{\bibfnamefont{W.~C.} \bibnamefont{Lai}},
  \bibinfo{author}{\bibfnamefont{N.}~\bibnamefont{McLean}},
  \bibinfo{author}{\bibfnamefont{A.}~\bibnamefont{Gansmuller}},
  \bibinfo{author}{\bibfnamefont{M.~A.} \bibnamefont{Verhoeven}},
  \bibinfo{author}{\bibfnamefont{G.~C.} \bibnamefont{Antonioli}},
  \bibinfo{author}{\bibfnamefont{M.}~\bibnamefont{Carravetta}},
  \bibinfo{author}{\bibfnamefont{L.}~\bibnamefont{Duma}},
  \bibinfo{author}{\bibfnamefont{P.~H.~M.} \bibnamefont{Bovee-Geurts}},
  \bibinfo{author}{\bibfnamefont{O.~G.} \bibnamefont{Johannessen}},
  \bibinfo{author}{\bibfnamefont{H.~J.~M.} \bibnamefont{de~Groot}},
  \bibnamefont{et~al.}, \bibinfo{journal}{J. Am. Chem. Soc.}
  \textbf{\bibinfo{volume}{128}}, \bibinfo{pages}{3878} (\bibinfo{year}{2006}).

\bibitem[{\citenamefont{Brown and Emsley}(2004)}]{brown04}
\bibinfo{author}{\bibfnamefont{S.~P.} \bibnamefont{Brown}} \bibnamefont{and}
  \bibinfo{author}{\bibfnamefont{L.}~\bibnamefont{Emsley}},
  \bibinfo{journal}{J. Magn. Reson.} \textbf{\bibinfo{volume}{171}},
  \bibinfo{pages}{43} (\bibinfo{year}{2004}).

\bibitem[{\citenamefont{Pham et~al.}(2007)\citenamefont{Pham, Griffin, Masiero,
  Leno, Gottarelli, Hodgkinson, Filip, and Brown}}]{pham07}
\bibinfo{author}{\bibfnamefont{T.~N.} \bibnamefont{Pham}},
  \bibinfo{author}{\bibfnamefont{J.~M.} \bibnamefont{Griffin}},
  \bibinfo{author}{\bibfnamefont{S.}~\bibnamefont{Masiero}},
  \bibinfo{author}{\bibfnamefont{S.}~\bibnamefont{Leno}},
  \bibinfo{author}{\bibfnamefont{G.}~\bibnamefont{Gottarelli}},
  \bibinfo{author}{\bibfnamefont{P.}~\bibnamefont{Hodgkinson}},
  \bibinfo{author}{\bibfnamefont{C.}~\bibnamefont{Filip}}, \bibnamefont{and}
  \bibinfo{author}{\bibfnamefont{S.~P.} \bibnamefont{Brown}},
  \bibinfo{journal}{Phys. Chem. Chem. Phys.} \textbf{\bibinfo{volume}{9}},
  \bibinfo{pages}{3416} (\bibinfo{year}{2007}).

\bibitem[{\citenamefont{Dingley and Grzesiek}(1998)}]{dingley98}
\bibinfo{author}{\bibfnamefont{A.~J.} \bibnamefont{Dingley}} \bibnamefont{and}
  \bibinfo{author}{\bibfnamefont{S.}~\bibnamefont{Grzesiek}},
  \bibinfo{journal}{J. Am. Chem. Soc} \textbf{\bibinfo{volume}{120}},
  \bibinfo{pages}{8293} (\bibinfo{year}{1998}).

\bibitem[{\citenamefont{Dingley et~al.}(2005)\citenamefont{Dingley, Peterson,
  Grzesiek, and Feigon}}]{dingley05}
\bibinfo{author}{\bibfnamefont{A.~J.} \bibnamefont{Dingley}},
  \bibinfo{author}{\bibfnamefont{R.~D.} \bibnamefont{Peterson}},
  \bibinfo{author}{\bibfnamefont{S.}~\bibnamefont{Grzesiek}}, \bibnamefont{and}
  \bibinfo{author}{\bibfnamefont{J.}~\bibnamefont{Feigon}},
  \bibinfo{journal}{J. Am. Chem. Soc.} \textbf{\bibinfo{volume}{127}},
  \bibinfo{pages}{14466} (\bibinfo{year}{2005}).

\bibitem[{\citenamefont{Vaara et~al.}(2002)\citenamefont{Vaara, Jokisaari,
  Wasylishen, and Bryce}}]{vaara02}
\bibinfo{author}{\bibfnamefont{J.}~\bibnamefont{Vaara}},
  \bibinfo{author}{\bibfnamefont{J.}~\bibnamefont{Jokisaari}},
  \bibinfo{author}{\bibfnamefont{R.~E.} \bibnamefont{Wasylishen}},
  \bibnamefont{and} \bibinfo{author}{\bibfnamefont{D.~L.} \bibnamefont{Bryce}},
  \bibinfo{journal}{Prog. Nucl. Magn. Reson. Spectrosc.}
  \textbf{\bibinfo{volume}{41}}, \bibinfo{pages}{233} (\bibinfo{year}{2002}).

\bibitem[{\citenamefont{Helgaker et~al.}(1999)\citenamefont{Helgaker,
  Jaszunski, and Ruud}}]{helgaker99}
\bibinfo{author}{\bibfnamefont{T.}~\bibnamefont{Helgaker}},
  \bibinfo{author}{\bibfnamefont{M.}~\bibnamefont{Jaszunski}},
  \bibnamefont{and} \bibinfo{author}{\bibfnamefont{K.}~\bibnamefont{Ruud}},
  \bibinfo{journal}{Chem. Rev.} \textbf{\bibinfo{volume}{99}},
  \bibinfo{pages}{293} (\bibinfo{year}{1999}).

\bibitem[{\citenamefont{Kaupp et~al.}(2004)\citenamefont{Kaupp, B\"uhl, and
  Malkin}}]{nmr_book}
\bibinfo{editor}{\bibfnamefont{M.}~\bibnamefont{Kaupp}},
  \bibinfo{editor}{\bibfnamefont{M.}~\bibnamefont{B\"uhl}}, \bibnamefont{and}
  \bibinfo{editor}{\bibfnamefont{V.~G.} \bibnamefont{Malkin}}, eds.,
  \emph{\bibinfo{title}{Calculation of NMR and EPR Parameters. Theory and
  Applications}} (\bibinfo{publisher}{Wiley VCH}, \bibinfo{address}{Weinheim},
  \bibinfo{year}{2004}).

\bibitem[{\citenamefont{Grzesiek et~al.}(2004)\citenamefont{Grzesiek, Cordier,
  Jaravine, and Barfield}}]{grzesiek04}
\bibinfo{author}{\bibfnamefont{S.}~\bibnamefont{Grzesiek}},
  \bibinfo{author}{\bibfnamefont{F.}~\bibnamefont{Cordier}},
  \bibinfo{author}{\bibfnamefont{V.}~\bibnamefont{Jaravine}}, \bibnamefont{and}
  \bibinfo{author}{\bibfnamefont{M.}~\bibnamefont{Barfield}},
  \bibinfo{journal}{Prog. Nucl. Magn. Reson. Spectrosc.}
  \textbf{\bibinfo{volume}{45}}, \bibinfo{pages}{275} (\bibinfo{year}{2004}).

\bibitem[{\citenamefont{Facelli and Grant}(1993)}]{facelli93}
\bibinfo{author}{\bibfnamefont{J.~C.} \bibnamefont{Facelli}} \bibnamefont{and}
  \bibinfo{author}{\bibfnamefont{D.~M.} \bibnamefont{Grant}},
  \bibinfo{journal}{Nature} \textbf{\bibinfo{volume}{365}},
  \bibinfo{pages}{325} (\bibinfo{year}{1993}).

\bibitem[{\citenamefont{Ochsenfeld et~al.}(2001)\citenamefont{Ochsenfeld,
  Brown, Schnell, Gauss, and Spiess}}]{ochsenfeld01}
\bibinfo{author}{\bibfnamefont{C.}~\bibnamefont{Ochsenfeld}},
  \bibinfo{author}{\bibfnamefont{S.~P.} \bibnamefont{Brown}},
  \bibinfo{author}{\bibfnamefont{I.}~\bibnamefont{Schnell}},
  \bibinfo{author}{\bibfnamefont{J.}~\bibnamefont{Gauss}}, \bibnamefont{and}
  \bibinfo{author}{\bibfnamefont{H.~W.} \bibnamefont{Spiess}},
  \bibinfo{journal}{J. Am. Chem. Soc.} \textbf{\bibinfo{volume}{123}},
  \bibinfo{pages}{2597} (\bibinfo{year}{2001}).

\bibitem[{\citenamefont{Salzmann et~al.}(1998)\citenamefont{Salzmann, Ziegler,
  Godbout, McMahon, Suslick, and Oldfield}}]{salzmann98}
\bibinfo{author}{\bibfnamefont{R.}~\bibnamefont{Salzmann}},
  \bibinfo{author}{\bibfnamefont{C.~J.} \bibnamefont{Ziegler}},
  \bibinfo{author}{\bibfnamefont{N.}~\bibnamefont{Godbout}},
  \bibinfo{author}{\bibfnamefont{M.~T.} \bibnamefont{McMahon}},
  \bibinfo{author}{\bibfnamefont{K.~S.} \bibnamefont{Suslick}},
  \bibnamefont{and} \bibinfo{author}{\bibfnamefont{E.}~\bibnamefont{Oldfield}},
  \bibinfo{journal}{J. Am. Chem. Soc.} \textbf{\bibinfo{volume}{120}},
  \bibinfo{pages}{11323} (\bibinfo{year}{1998}).

\bibitem[{\citenamefont{Pickard and Mauri}(2001)}]{pickard01}
\bibinfo{author}{\bibfnamefont{C.~J.} \bibnamefont{Pickard}} \bibnamefont{and}
  \bibinfo{author}{\bibfnamefont{F.}~\bibnamefont{Mauri}},
  \bibinfo{journal}{Phys. Rev. B} \textbf{\bibinfo{volume}{63}},
  \bibinfo{pages}{245101} (\bibinfo{year}{2001}).

\bibitem[{\citenamefont{Payne et~al.}(1992)\citenamefont{Payne, Teter, Allen,
  Arias, and Joannopoulos}}]{payne1}
\bibinfo{author}{\bibfnamefont{M.~C.} \bibnamefont{Payne}},
  \bibinfo{author}{\bibfnamefont{M.~P.} \bibnamefont{Teter}},
  \bibinfo{author}{\bibfnamefont{D.~C.} \bibnamefont{Allen}},
  \bibinfo{author}{\bibfnamefont{T.~A.} \bibnamefont{Arias}}, \bibnamefont{and}
  \bibinfo{author}{\bibfnamefont{J.~D.} \bibnamefont{Joannopoulos}},
  \bibinfo{journal}{Rev. Mod. Phys.} \textbf{\bibinfo{volume}{64}},
  \bibinfo{pages}{1045} (\bibinfo{year}{1992}).

\bibitem[{\citenamefont{Ashbrook et~al.}(2006)\citenamefont{Ashbrook, Polles,
  Gautier, Pickard, and Walton}}]{ashbrook06}
\bibinfo{author}{\bibfnamefont{S.~E.} \bibnamefont{Ashbrook}},
  \bibinfo{author}{\bibfnamefont{L.~L.} \bibnamefont{Polles}},
  \bibinfo{author}{\bibfnamefont{R.}~\bibnamefont{Gautier}},
  \bibinfo{author}{\bibfnamefont{C.~J.} \bibnamefont{Pickard}},
  \bibnamefont{and} \bibinfo{author}{\bibfnamefont{R.~I.}
  \bibnamefont{Walton}}, \bibinfo{journal}{Phys. Chem. Chem. Phys.}
  \textbf{\bibinfo{volume}{8}}, \bibinfo{pages}{3423} (\bibinfo{year}{2006}).

\bibitem[{\citenamefont{Farnan et~al.}(2003)\citenamefont{Farnan, Balan,
  Pickard, and Mauri}}]{farnan03}
\bibinfo{author}{\bibfnamefont{I.}~\bibnamefont{Farnan}},
  \bibinfo{author}{\bibfnamefont{E.}~\bibnamefont{Balan}},
  \bibinfo{author}{\bibfnamefont{C.~J.} \bibnamefont{Pickard}},
  \bibnamefont{and} \bibinfo{author}{\bibfnamefont{F.}~\bibnamefont{Mauri}},
  \bibinfo{journal}{Am. Miner.} \textbf{\bibinfo{volume}{88}},
  \bibinfo{pages}{1663} (\bibinfo{year}{2003}).

\bibitem[{\citenamefont{Profeta et~al.}(2003)\citenamefont{Profeta, Mauri, and
  Pickard}}]{profeta03}
\bibinfo{author}{\bibfnamefont{M.}~\bibnamefont{Profeta}},
  \bibinfo{author}{\bibfnamefont{F.}~\bibnamefont{Mauri}}, \bibnamefont{and}
  \bibinfo{author}{\bibfnamefont{C.~J.} \bibnamefont{Pickard}},
  \bibinfo{journal}{J. Am. Chem. Soc.} \textbf{\bibinfo{volume}{125}},
  \bibinfo{pages}{541} (\bibinfo{year}{2003}).

\bibitem[{\citenamefont{Benoit et~al.}(2005)\citenamefont{Benoit, Profeta,
  Mauri, Pickard, and Tuckerman}}]{benoit05}
\bibinfo{author}{\bibfnamefont{M.}~\bibnamefont{Benoit}},
  \bibinfo{author}{\bibfnamefont{M.}~\bibnamefont{Profeta}},
  \bibinfo{author}{\bibfnamefont{F.}~\bibnamefont{Mauri}},
  \bibinfo{author}{\bibfnamefont{C.~J.} \bibnamefont{Pickard}},
  \bibnamefont{and} \bibinfo{author}{\bibfnamefont{M.~E.}
  \bibnamefont{Tuckerman}}, \bibinfo{journal}{J. Phys. Chem. B}
  \textbf{\bibinfo{volume}{109}}, \bibinfo{pages}{6052} (\bibinfo{year}{2005}).

\bibitem[{\citenamefont{Charpentier et~al.}(2004)\citenamefont{Charpentier,
  Ispas, Profeta, Mauri, and Pickard}}]{charpentier04}
\bibinfo{author}{\bibfnamefont{T.}~\bibnamefont{Charpentier}},
  \bibinfo{author}{\bibfnamefont{S.}~\bibnamefont{Ispas}},
  \bibinfo{author}{\bibfnamefont{M.}~\bibnamefont{Profeta}},
  \bibinfo{author}{\bibfnamefont{F.}~\bibnamefont{Mauri}}, \bibnamefont{and}
  \bibinfo{author}{\bibfnamefont{C.~J.} \bibnamefont{Pickard}},
  \bibinfo{journal}{J. Phys. Chem. B} \textbf{\bibinfo{volume}{108}},
  \bibinfo{pages}{4147} (\bibinfo{year}{2004}).

\bibitem[{\citenamefont{Yates et~al.}(2004)\citenamefont{Yates, Pickard, Payne,
  Dupree, Profeta, and Mauri}}]{yates04}
\bibinfo{author}{\bibfnamefont{J.~R.} \bibnamefont{Yates}},
  \bibinfo{author}{\bibfnamefont{C.~J.} \bibnamefont{Pickard}},
  \bibinfo{author}{\bibfnamefont{M.~C.} \bibnamefont{Payne}},
  \bibinfo{author}{\bibfnamefont{R.}~\bibnamefont{Dupree}},
  \bibinfo{author}{\bibfnamefont{M.}~\bibnamefont{Profeta}}, \bibnamefont{and}
  \bibinfo{author}{\bibfnamefont{F.}~\bibnamefont{Mauri}}, \bibinfo{journal}{J.
  Phys. Chem. A} \textbf{\bibinfo{volume}{108}}, \bibinfo{pages}{6032}
  (\bibinfo{year}{2004}).

\bibitem[{\citenamefont{Gervais et~al.}(2005)\citenamefont{Gervais, Dupree,
  Pike, Bonhomme, Profeta, Pickard, and Mauri}}]{gervais05}
\bibinfo{author}{\bibfnamefont{C.}~\bibnamefont{Gervais}},
  \bibinfo{author}{\bibfnamefont{R.}~\bibnamefont{Dupree}},
  \bibinfo{author}{\bibfnamefont{K.~J.} \bibnamefont{Pike}},
  \bibinfo{author}{\bibfnamefont{C.}~\bibnamefont{Bonhomme}},
  \bibinfo{author}{\bibfnamefont{M.}~\bibnamefont{Profeta}},
  \bibinfo{author}{\bibfnamefont{C.~J.} \bibnamefont{Pickard}},
  \bibnamefont{and} \bibinfo{author}{\bibfnamefont{F.}~\bibnamefont{Mauri}},
  \bibinfo{journal}{J. Phys. Chem. A} \textbf{\bibinfo{volume}{109}},
  \bibinfo{pages}{6960} (\bibinfo{year}{2005}).

\bibitem[{\citenamefont{Yates et~al.}(2005)\citenamefont{Yates, Pham, Pickard,
  Mauri, Amado, Gil, and Brown}}]{yates05-malt}
\bibinfo{author}{\bibfnamefont{J.~R.} \bibnamefont{Yates}},
  \bibinfo{author}{\bibfnamefont{T.~N.} \bibnamefont{Pham}},
  \bibinfo{author}{\bibfnamefont{C.~J.} \bibnamefont{Pickard}},
  \bibinfo{author}{\bibfnamefont{F.}~\bibnamefont{Mauri}},
  \bibinfo{author}{\bibfnamefont{A.~M.} \bibnamefont{Amado}},
  \bibinfo{author}{\bibfnamefont{A.~M.} \bibnamefont{Gil}}, \bibnamefont{and}
  \bibinfo{author}{\bibfnamefont{S.~P.} \bibnamefont{Brown}},
  \bibinfo{journal}{J. Am. Chem. Soc.} \textbf{\bibinfo{volume}{127}},
  \bibinfo{pages}{10216} (\bibinfo{year}{2005}).

\bibitem[{\citenamefont{Bl\"{o}chl}(1994)}]{blochl1}
\bibinfo{author}{\bibfnamefont{P.~E.} \bibnamefont{Bl\"{o}chl}},
  \bibinfo{journal}{Phys. Rev. B} \textbf{\bibinfo{volume}{50}},
  \bibinfo{pages}{17 953} (\bibinfo{year}{1994}).

\bibitem[{\citenamefont{Ramsey and Purcell}(1952)}]{ramsey52}
\bibinfo{author}{\bibfnamefont{N.~F.} \bibnamefont{Ramsey}} \bibnamefont{and}
  \bibinfo{author}{\bibfnamefont{E.~M.} \bibnamefont{Purcell}},
  \bibinfo{journal}{Phys. Rev.} \textbf{\bibinfo{volume}{85}},
  \bibinfo{pages}{143} (\bibinfo{year}{1952}).

\bibitem[{\citenamefont{Ramsey}(1953)}]{ramsey53}
\bibinfo{author}{\bibfnamefont{N.~F.} \bibnamefont{Ramsey}},
  \bibinfo{journal}{Phys. Rev.} \textbf{\bibinfo{volume}{91}},
  \bibinfo{pages}{303} (\bibinfo{year}{1953}).

\bibitem[{\citenamefont{Marquez et~al.}(2001)\citenamefont{Marquez, Gerwick,
  and Willianson}}]{marquez01}
\bibinfo{author}{\bibfnamefont{B.~L.} \bibnamefont{Marquez}},
  \bibinfo{author}{\bibfnamefont{W.~H.} \bibnamefont{Gerwick}},
  \bibnamefont{and} \bibinfo{author}{\bibfnamefont{R.~T.}
  \bibnamefont{Willianson}}, \bibinfo{journal}{Magn. Reson. Chem.}
  \textbf{\bibinfo{volume}{39}}, \bibinfo{pages}{449} (\bibinfo{year}{2001}).

\bibitem[{\citenamefont{Edden and Keeler}(2004)}]{edden04}
\bibinfo{author}{\bibfnamefont{R.~A.~E.} \bibnamefont{Edden}} \bibnamefont{and}
  \bibinfo{author}{\bibfnamefont{J.}~\bibnamefont{Keeler}},
  \bibinfo{journal}{J. Mag. Res.} \textbf{\bibinfo{volume}{166}},
  \bibinfo{pages}{53} (\bibinfo{year}{2004}).

\bibitem[{\citenamefont{van~de Walle and Bl\"ochl}(1993)}]{walle1}
\bibinfo{author}{\bibfnamefont{C.~G.} \bibnamefont{van~de Walle}}
  \bibnamefont{and} \bibinfo{author}{\bibfnamefont{P.~E.}
  \bibnamefont{Bl\"ochl}}, \bibinfo{journal}{Phys. Rev. B}
  \textbf{\bibinfo{volume}{47}}, \bibinfo{pages}{4244} (\bibinfo{year}{1993}).

\bibitem[{\citenamefont{Petrilli et~al.}(1998)\citenamefont{Petrilli,
  Bl\"{o}chl, Blaha, and Schwarz}}]{efg1}
\bibinfo{author}{\bibfnamefont{H.~M.} \bibnamefont{Petrilli}},
  \bibinfo{author}{\bibfnamefont{P.~E.} \bibnamefont{Bl\"{o}chl}},
  \bibinfo{author}{\bibfnamefont{P.}~\bibnamefont{Blaha}}, \bibnamefont{and}
  \bibinfo{author}{\bibfnamefont{K.}~\bibnamefont{Schwarz}},
  \bibinfo{journal}{Phys. Rev. B} \textbf{\bibinfo{volume}{57}},
  \bibinfo{pages}{14690} (\bibinfo{year}{1998}).

\bibitem[{\citenamefont{Pickard and Payne}(1997)}]{eels1}
\bibinfo{author}{\bibfnamefont{C.~J.} \bibnamefont{Pickard}} \bibnamefont{and}
  \bibinfo{author}{\bibfnamefont{M.~C.} \bibnamefont{Payne}},
  \bibinfo{journal}{Inst. Phys. Conf. Ser.} \textbf{\bibinfo{volume}{153}},
  \bibinfo{pages}{179} (\bibinfo{year}{1997}).

\bibitem[{\citenamefont{Gonze}(1995)}]{gonze97}
\bibinfo{author}{\bibfnamefont{X.}~\bibnamefont{Gonze}},
  \bibinfo{journal}{Phys. Rev. B} \textbf{\bibinfo{volume}{55}},
  \bibinfo{pages}{10337} (\bibinfo{year}{1995}).

\bibitem[{\citenamefont{Clark et~al.}(2005)\citenamefont{Clark, Segall,
  Pickard, Hasnip, Probert, Refson, and Payne}}]{clark05}
\bibinfo{author}{\bibfnamefont{S.~J.} \bibnamefont{Clark}},
  \bibinfo{author}{\bibfnamefont{M.~D.} \bibnamefont{Segall}},
  \bibinfo{author}{\bibfnamefont{C.~J.} \bibnamefont{Pickard}},
  \bibinfo{author}{\bibfnamefont{P.~J.} \bibnamefont{Hasnip}},
  \bibinfo{author}{\bibfnamefont{M.~J.} \bibnamefont{Probert}},
  \bibinfo{author}{\bibfnamefont{K.}~\bibnamefont{Refson}}, \bibnamefont{and}
  \bibinfo{author}{\bibfnamefont{M.~C.} \bibnamefont{Payne}},
  \bibinfo{journal}{Z. Kristall.} \textbf{\bibinfo{volume}{220}},
  \bibinfo{pages}{567} (\bibinfo{year}{2005}).

\bibitem[{\citenamefont{Troullier and Martins}(1991)}]{troullier1}
\bibinfo{author}{\bibfnamefont{N.}~\bibnamefont{Troullier}} \bibnamefont{and}
  \bibinfo{author}{\bibfnamefont{J.~L.} \bibnamefont{Martins}},
  \bibinfo{journal}{Phys. Rev. B} \textbf{\bibinfo{volume}{43}},
  \bibinfo{pages}{1993} (\bibinfo{year}{1991}).

\bibitem[{not({\natexlab{a}})}]{notemol}
\bibinfo{note}{C$_{2}$H$_{2}$, C$_{2}$H$_{4}$, C$_{2}$H$_{6}$, C$_{6}$H$_{6}$,
  HCN, HNC, CH$_{3}$CN, CH$_{3}$NC, CH$_{3}$F, CHF$_{3}$, OF$_{2}$,
  CH$_{3}$SiH$_{3}$, HF, HCl, H$_{2}$O, CH$_{4}$, SiH$_{4}$}.

\bibitem[{\citenamefont{Lantto et~al.}(2002)\citenamefont{Lantto, Vaara, and
  Helgaker}}]{lantto02}
\bibinfo{author}{\bibfnamefont{P.}~\bibnamefont{Lantto}},
  \bibinfo{author}{\bibfnamefont{J.}~\bibnamefont{Vaara}}, \bibnamefont{and}
  \bibinfo{author}{\bibfnamefont{T.}~\bibnamefont{Helgaker}},
  \bibinfo{journal}{J. Chem. Phys} \textbf{\bibinfo{volume}{117}},
  \bibinfo{pages}{5998} (\bibinfo{year}{2002}).

\bibitem[{\citenamefont{Perdew et~al.}(1996)\citenamefont{Perdew, Burke, and
  Ernzerhof}}]{perdew1}
\bibinfo{author}{\bibfnamefont{J.~P.} \bibnamefont{Perdew}},
  \bibinfo{author}{\bibfnamefont{K.}~\bibnamefont{Burke}}, \bibnamefont{and}
  \bibinfo{author}{\bibfnamefont{M.}~\bibnamefont{Ernzerhof}},
  \bibinfo{journal}{Phys. Rev. Lett.} \textbf{\bibinfo{volume}{77}},
  \bibinfo{pages}{3865} (\bibinfo{year}{1996}).

\bibitem[{\citenamefont{Kaski et~al.}(1996)\citenamefont{Kaski, Vaara, and
  Jokisaari}}]{kaski96}
\bibinfo{author}{\bibfnamefont{J.}~\bibnamefont{Kaski}},
  \bibinfo{author}{\bibfnamefont{J.}~\bibnamefont{Vaara}}, \bibnamefont{and}
  \bibinfo{author}{\bibfnamefont{J.}~\bibnamefont{Jokisaari}},
  \bibinfo{journal}{J. Am. Chem. Soc.} \textbf{\bibinfo{volume}{118}},
  \bibinfo{pages}{8879} (\bibinfo{year}{1996}).

\bibitem[{\citenamefont{Fletcher et~al.}(1996)\citenamefont{Fletcher,
  McMeeking, and Parkin}}]{cds}
\bibinfo{author}{\bibfnamefont{D.}~\bibnamefont{Fletcher}},
  \bibinfo{author}{\bibfnamefont{R.}~\bibnamefont{McMeeking}},
  \bibnamefont{and} \bibinfo{author}{\bibfnamefont{D.}~\bibnamefont{Parkin}},
  \bibinfo{journal}{J. Chem. Inf. Comput. Sci.} \textbf{\bibinfo{volume}{36}},
  \bibinfo{pages}{746} (\bibinfo{year}{1996}).

\bibitem[{not({\natexlab{b}})}]{noteref}
\bibinfo{note}{The reference shieldings required to convert the calculated
  shieldings into chemical shifts were obtained from
  Ref.~\onlinecite{profeta03} for Si and Ref.~\onlinecite{profeta04} for P.}

\bibitem[{\citenamefont{Profeta et~al.}(2004)\citenamefont{Profeta, Benoit,
  Mauri, and Pickard}}]{profeta04}
\bibinfo{author}{\bibfnamefont{M.}~\bibnamefont{Profeta}},
  \bibinfo{author}{\bibfnamefont{M.}~\bibnamefont{Benoit}},
  \bibinfo{author}{\bibfnamefont{F.}~\bibnamefont{Mauri}}, \bibnamefont{and}
  \bibinfo{author}{\bibfnamefont{C.~J.} \bibnamefont{Pickard}},
  \bibinfo{journal}{J. Am. Chem. Soc.} \textbf{\bibinfo{volume}{126}},
  \bibinfo{pages}{12628} (\bibinfo{year}{2004}).

\end{thebibliography}
\end{document}